\documentclass[runningheads]{llncs}
\usepackage[T1]{fontenc}
\usepackage{graphicx}
\usepackage{hyperref}
\usepackage{color}

\usepackage{orcidlink}
\usepackage[capitalize]{cleveref}

\newif\ifanonymized
\anonymizedfalse  

\begin{document}
\title{Composable Building Blocks for Controllable and Transparent Interactive AI Systems}
\titlerunning{Building Blocks for Controllable and Transparent Interactive AI Systems}
%
\ifanonymized
    \author{Anonymized}
\else
    \author{
        Sebe Vanbrabant\,\orcidlink{0009-0001-7996-6048} \and
        Gustavo Rovelo Ruiz\,\orcidlink{0000-0001-7580-8950} \and
        Davy Vanacken\,\orcidlink{0000-0001-8436-5119}
    }
\fi
\ifanonymized
    \authorrunning{Anonymized}
\else
    \authorrunning{S. Vanbrabant et al.}
\fi
%
\ifanonymized
    \institute{Anonymized}
\else
    \institute{Hasselt University - Flanders Make, Digital Future Lab, Diepenbeek, Belgium
        \email{\href{mailto:sebe.vanbrabant@uhasselt.be}{sebe.vanbrabant@uhasselt.be}}
        \email{\href{mailto:gustavo.roveloruiz@uhasselt.be}{gustavo.roveloruiz@uhasselt.be}}
        \email{\href{mailto:davy.vanacken@uhasselt.be}{davy.vanacken@uhasselt.be}}
    }
\fi
\maketitle              
\begin{abstract}
While the increased integration of AI technologies into interactive systems enables them to solve an equally increasing number of tasks, the black box problem of AI models continues to spread throughout the interactive system as a whole. Explainable AI (XAI) techniques can make AI models more accessible by employing post-hoc methods or transitioning to inherently interpretable models. While this makes individual AI models clearer, the overarching system architecture remains opaque. To this end, we propose an approach to represent interactive systems as sequences of structural building blocks, such as AI models and control mechanisms grounded in the literature. These can then be explained through accompanying visual building blocks, such as XAI techniques. The flow and APIs of the structural building blocks form an explicit overview of the system. This serves as a communication basis for both humans and automated agents like LLMs, aligning human and machine interpretability of AI models. We discuss a selection of building blocks and concretize our flow-based approach in an architecture and accompanying prototype interactive system.

\end{abstract}
\section{Introduction}
Artificial intelligence (AI) is becoming increasingly integrated into various interactive systems, with different challenges arising depending on the AI technologies used and the type of user interactions offered by these systems~\cite{dix2023engineering}.
The increasing complexity of AI models, from interpretable decision trees to opaque Deep Neural Networks (DNNs) and Large Language Models (LLMs), has led to a decline in their transparency~\cite{arrieta2020explainable,luo2024understanding}. These models are often black boxes, producing results without explanations, justifications, or indications of uncertainties~\cite{von2021transparency}.
The field of eXplainable AI (XAI) addresses these challenges by complementing AI predictions with explanations~\cite{gunning2019xai,xu2023xair}. Machine learning (ML) workflows can be made more transparent in two ways: either by using white-box models that offer inherent interpretability, like decision trees, or by leveraging post-hoc explanations (e.g., LIME~\cite{ribeiro2016should} and SHAP~\cite{lundberg2017unified}) to try to explain the internal workings of black box models, such as neural networks.

We view explainability techniques like LIME, SHAP and the What-If tool~\cite{wexler2019if} as visual \textit{building blocks}. They address AI models' transparency by answering Why, Why-not, and What-if. However, no widespread building blocks exist that support users to control their AI models in the same way that LIME and SHAP address transparency. Current visual approaches can explicate AI behavior by interacting with the model (e.g., allow the user to change model inputs~\cite{wexler2019if,he2025conversational}) or through overviews of its internals (e.g., the structure of a neural network~\cite{calo2025deepflow}).

While standardized approaches exist for interpreting model behavior, they are not necessarily applicable to the interactive system in which they are embedded. Kulesza et al.~\cite{kulesza2012tell} found that the quality of a user’s mental model directly correlates to their ability to control the underlying system as desired. This also applies to LLMs, which require careful prompting and the right (amount of) information to address user queries accurately. Approaches like Tool-Augmented Language Models (TALMs)~\cite{parisi2022talm} and Anthropic's recent Model Context Protocol (MCP)~\cite{mcp} enable LLMs to invoke code subroutines, facilitating their integration into interactive systems.

We envision an approach that simultaneously empowers users and automated agents to understand and control AI models. This involves extending XAI techniques and subroutine-based tools beyond \textit{model-level} explanations. We instead look to support transparency and control in \textit{system-level} AI workflows through structural and visual building blocks. For example, an AI model (structural building block) can be explained through LIME, SHAP, and WhatIf (visual building blocks), and further controlled through structured building blocks that, for instance, override unintended decisions~\cite{kieseberg2023controllable} or give per-instance feedback~\cite{kulesza2015principles}. To combine these blocks into one approach, we draw inspiration from neuro-symbolic AI (NSAI), which integrates neural and symbolic approaches to combine their strengths while circumventing their inherent weaknesses~\cite{wang2025towards}. By incorporating techniques to explain one AI model into conceptual systems using structural building blocks, we aim to clarify interactive AI systems for humans and enable automated tools and agents, such as LLMs, to audit them using a shared knowledge base, aligning human and machine interpretation of AI models.

\section{Related Work}
Looking at the nine stages of the ML workflow, two major stages are evident: one data-oriented stage, involving data preparation, and one model-oriented stage, involving model (re)training and deployment~\cite{amershi2019software}. For supervised ML, this results in a model that can predict new outputs from new inputs by leveraging its internal learning process. We can, thus, conceptually, view a trained model as a pipeline that transforms inputs into outputs through a model. These pipelines can be chained to make system behavior more advanced and fit for a task, which is the case for interactive systems embedding AI technologies.
\textit{Symbolic} AI, such as decision rules, excels at structured reasoning and provides high inherent explainability and interpretability~\cite{wang2025towards}. However, symbolic approaches are less trainable and more error-prone in unfamiliar situations. In contrast, \textit{connectionist} techniques like neural networks excel at training by discovering and learning patterns from data, yet remain black boxes that require large datasets for effective training.

NSAI combines trainability and interpretability by using neural approaches to learn from experience and applying symbolic reasoning to draw conclusions from that knowledge~\cite{wang2025towards}. Type 2 NSAI, as described by Kautz~\cite{kautz2022third}, considers connectionist models as neural module subroutines within a symbolic problem-solving system. TALMs are a recent example of type 2 NSAI systems. Systems like ViperGPT~\cite{suris2023vipergpt} and Chameleon~\cite{lu2023chameleon} combine LLMs as neural subroutines within a symbolic tool usage framework. TALMs query tools rather than generating the answer directly, which is helpful for mathematical operations or to interface with external APIs.


The strengths of LLMs for XAI are evident in x-[plAIn]~\cite{mavrepis2024xai} and SHAPstories~\cite{martens2025tell}, which generate audience-specific summaries of XAI methods tailored to users’ knowledge and interests, improving accessibility and decision-making. These approaches, however, do not offer capabilities other than those of the XAI methods. ECHO~\cite{vanbrabant2025echo}, a conversational approach to XAI, tackles this with a TALM using generated tools for explicating system-specific behavior complemented with predefined tools that address various explanation types and XAI methods. These approaches are all purely textual, however, and could be extended to intelligible interfaces. For instance, visualizations like those in TimberTrek~\cite{wang2022timbertrek} and AI-Spectra~\cite{eerlings2024ai} can be enhanced by integrating conversational interfaces to help users understand and select the right models for their needs. An explainer offering recommendations and explanatory insights can make the process more accessible and interactive.

Aside from visualizing model analysis, other tools offer ways to build models visually using flow-based graph-like visualizations. To visualize models during development, DeepGraph~\cite{hu2018deepgraph} constructs the data flow graph representation of the architecture from the DNN source code and automatically synchronizes it with its graph representation. To enable users to build deep learning models visually, DeepFlow~\cite{calo2025deepflow} uses a flow-based visual programming tool, realizing a no-code approach to building neural networks while viewing models as sequences of learnable functions. Existing approaches help visualize complex architectures and democratize AI development, but primarily focus on the AI development process rather than the model's role within the encapsulating AI system.

\section{Building Blocks for Intelligibility and Control}
While current approaches to explainability typically interrogate AI models by probing parts of the \textit{input $\rightarrow$ model $\rightarrow$ output} pipeline through the Predict method, we propose expanding this conventional pipeline through structural and visual building blocks that also allow AI models to be visualized and controlled. By considering interactive systems embedding AI as type 2 NSAI systems consisting of specific components, their decision process can be represented through structural building blocks, which can be explained through visual building blocks. This gives both humans and AI agents a structured and shared knowledge base of (complex) system architectures through accessible building blocks.

\subsection{Visual Building Blocks}

\subsubsection{Intelligiblity}
We initially considered the commonly used explanations of `Why', `Why-not', and `What-if'~\cite{mohseni2021multidisciplinary}. Why and Why-not can be addressed by visual building blocks encompassing \textbf{LIME} and \textbf{SHAP}, which explain AI behavior using feature importance, which is commonly used to address Why and Why-not explanations. For What-if questions, we use a visual building block displaying all the predict method's input parameters and a corresponding output value, similar to the approach of He et al.~\cite{he2025conversational}.

\subsection{Structural Building Blocks}

\subsubsection{Control}
One way to honor user feedback through a structural building block would be to allow users to re-label instances and retrain the model accordingly~\cite{kulesza2015principles}. We draw further inspiration from the work on controllable AI by Kieseberg et al.~\cite{kieseberg2023controllable}. Their five methods for managing control loss map to structural building blocks in our interactive NSAI pipeline. For non-autonomous AI systems, \textbf{DivineRuleGuard} ensures ethical compliance by overriding harmful or unethical decisions before they are acted upon as a postprocessing step for model output. Conversely, \textbf{NonGoalFilter} acts as a pre-processor, rejecting inputs that do not align with intended behaviors or intentions. \textbf{ShutdownTrigger} functions as an emergency stop to disable autonomous AI systems at any point. \textbf{BiasInjector} strategically influences decision-making by embedding predefined biases to guide the model toward preferred outcomes. Lastly, \textbf{LogicBomb} operates as a self-monitoring fail-safe, resetting or shutting down the AI if it ever attempts to produce an outcome that breaches ethical or operational boundaries.


\subsubsection{Execution Flow}
Aside from controlling the individual NSAI components, we also envision components for visualizing and controlling conditional execution flows between components. This is useful in scenarios where multiple AI models are used together, such as in the context of ensemble learning~\cite{opitz1999popular} or model multiplicity~\cite{wang2022timbertrek,eerlings2024ai}, techniques commonly used in high-stakes interactive systems. Specifically, we envision a \textbf{Splitter} and \textbf{Aggregator}, where the Splitter indicates the dataset being distributed to different AI models. The Aggregator then displays how the final output is produced/aggregated from these models. For model multiplicity, Chernoff bots could be the visual building block for each individual model; its dashboard can be linked to the aggregator. 

\section{Auditable 5-Layer Architecture for Transparent and Controllable Interactive AI Systems}
We apply our proposed 5-layer architecture to an example heart disease prediction ensemble in \cref{fig:architecture}. The vertical pipeline was loosely based on the XAI system depicted by Mohseni et al.~\cite{mohseni2021multidisciplinary}, modified to support structural and visual building blocks. We define the following layers:

\begin{figure}[h]
  \centering
  \includegraphics[width=1\linewidth]{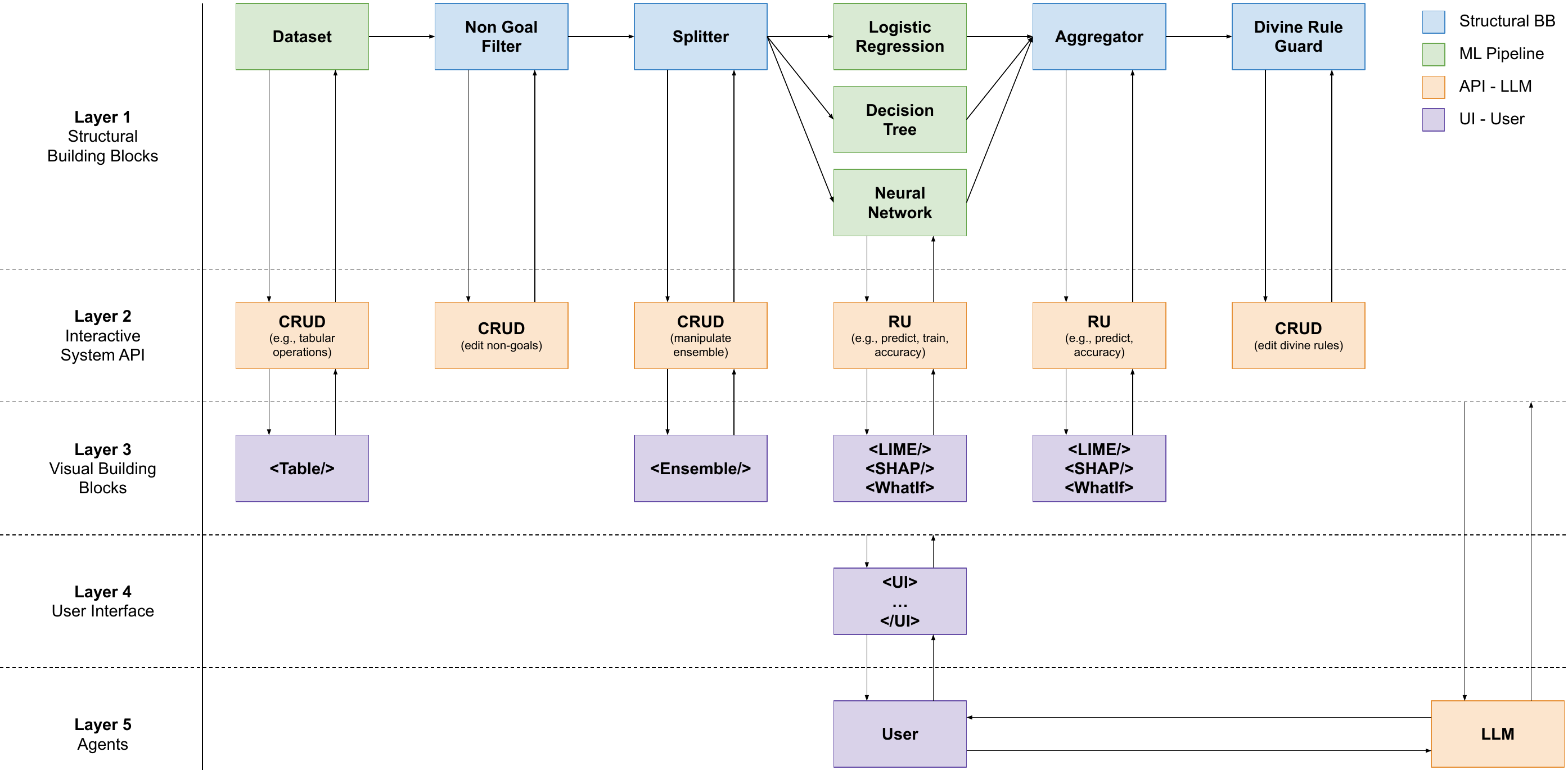}
  \caption{An example heart disease prediction ensemble to illustrate our proposed architecture. Layer~1 shows the structural building blocks, consisting of our building blocks (in blue) integrated into the ML pipeline (in green). Layer~2 converts the structural building blocks into a callable API, usable by the visual building blocks in layer~3, assembled into the user interface of layer~4. The API is also accessible by the LLM of layer~5 so that both agents use common knowledge.}
  \label{fig:architecture}
\end{figure}


\begin{description}
    \item[Layer~1: Structural Building Blocks]{Structural building blocks convey the (conceptual) architecture of the interactive system at a glance. Each block is mapped to a part of the system's source code, completely specified by the developer through function decorators. Since this representation is purely conceptual, the developer can choose what (parts of) the system to expose and how to communicate the pipeline.}

    \item[Layer~2: Interactive System API]{Developers can write their code as usual, and link it to conceptual structural building blocks through developer-defined methods. Each block's REST API is automatically generated from the structural building blocks' definition and its decorator of layer~1.}

    \item[Layer~3: Visual Building Blocks]{Visual building blocks interact with the structural building blocks of layer~1 through the API of layer~2. For instance, a structural building block of an AI model can have its behavior explained through LIME, for which data is acquired over the REST API.}

    \item[Layer~4: User Interface]{
    Structural and visual building blocks are combined into an interface where they can be explored, interrogated, and controlled. Currently, visual building blocks are assembled into one coherent UI; future research directions include LLM-powered layouts~\cite{brie2023evaluating}.}

    \item[Layer~5: Agents]{The final layer of the architecture comprises agents that interact with the building blocks. These can be users interacting with the UI and its visual building blocks, or an automated LLM agent interacting with the shared REST API of layer~2. This API can then be integrated as tools for a TALM such as ECHO~\cite{vanbrabant2025echo}. Both agents have access to the same information, and users can interact with the LLM to ask about system behavior.}
\end{description}

\begin{figure}[h]
  \centering
  \includegraphics[width=1\linewidth]{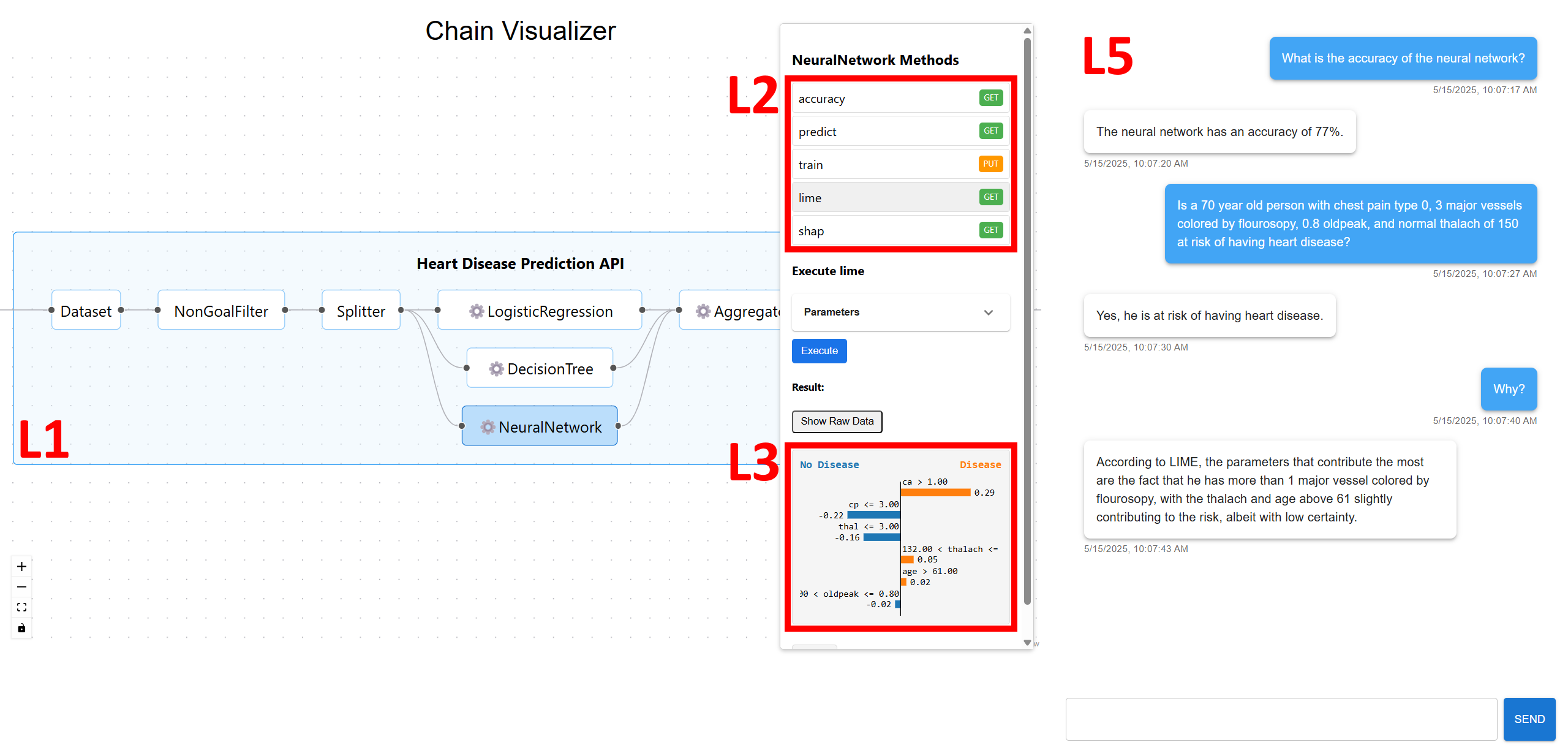}
  \caption{Prototype of our approach using the heart disease prediction ensemble. The UI (layer~4) shows each structural building block (layer~1) influencing predictions, exposing system behavior to both users and automated agents (layer~5) through visual building blocks (layer~3) and an API (layer~2), respectively.}
  \label{fig:usecase}
\end{figure}

\section{Conclusion}
Rising AI complexity has led to an increase in challenges regarding explaining and controlling AI behavior. These challenges propagate from the individual model to the system that embeds it, making the entire interactive system opaque to users. We proposed a preliminary architecture for making interactive systems more accessible by explicitly conveying their conceptual model through an API. This enables both users and LLMs to access information related to system behavior, aligning human and machine interpretability of AI models. Future research directions include applying our architecture to more elaborate NSAI applications and use cases involving model multiplicity, such as integrating the AI-Spectra dashboard and its Chernoff bots as visual building blocks~\cite{eerlings2024ai}. Furthermore, it would be interesting to explore and integrate user-specific, personalized, and dynamic UIs into the textual conversations that adapt to individual user needs.

\ifanonymized
\else
    \begin{credits}
    \subsubsection{\ackname}
    This work was funded by the Special Research Fund (BOF) of Hasselt University, BOF23OWB31.
    
    \subsubsection{\discintname}
    The authors have no competing interests to declare that
    are relevant to the content of this article.
    \end{credits}
\fi
%
%
%
\bibliographystyle{splncs04}
\bibliography{references}
\end{document}